# A Theoretical Approach for a Novel Model to Realizing Empathy

Marialejandra García Corretjer

*GTM-Grup de recerca en Tecnologies Mèdia, La Salle-Universitat Ramon Llull, Barcelona, Spain*

Carrer Quatre Camins 30, Barcelona, Spain, 08022, ma.garcia@salle.url.edu

The focus of Marialejandra's current line of research is the result of a diverse industry & academic experiences converging into one challenge. Throughout her career, after two Masters degrees (Multimedia Project Management and Research for Design & Innovation) and a Bachelor degree in Industrial Design & Fine Arts, she was always interested in the relationships between people and their objects, as a form of communication that reveals preferences, expectations, and personalities. Now, she studies the impact of this relationship between humans and smart objects within the Human Computer Interactions discipline, from the perspective of the process of Realizing Empathy.

David Miralles

*GTM-Grup de recerca en Tecnologies Mèdia, La Salle-Universitat Ramon Llull, Barcelona, Spain*

Carrer Quatre Camins 30, Barcelona, Spain, david.miralles@salle.url.edu

He holds a degree on Theoretical Physics of the University of Barcelona (1995). From 1996 to 2001 he worked at the Fundamental Physics Department at the same university. In 2001 he obtained a PhD on Mathematical Physics. From 2001 to 2007 he worked at the Department of Communication and Signal Theory of Universitat Ramon Llull. Nowadays, he is member of the Grup de recerca en Tecnologies Media at La Salle-Universitat Ramon Llull.

Raquel Ros

*GTM-Grup de recerca en Tecnologies Mèdia, La Salle-Universitat Ramon Llull, Barcelona, Spain*

Carrer Quatre Camins 30, Barcelona, Spain, raquel.ros@salle.url.edu

# A Theoretical Approach for a Novel Model to Realizing Empathy

## 1. Introduction

Recent HCI literary works within affective computing, social robots, and person-AI cooperation, focus on topics related to: fostering a relationship between people and their technology [10], creating tools to engage them in collaboration to empower their knowledge and creativity [38], and the role of empathy in these scenarios [45]. As empathy continues to play a central role in how people relate with each other & becomes the basis for how people perceive their objects, and contexts the term itself has been a topic of constant discussion and debate across fields of study. Beyond understanding the nature of empathy between people, their relationship with advanced technology and smart objects also become the subject of empathy related observations and scrutiny [29]. One of the main concerns among authors that approach empathy as a component of human-machine experiences and relationships is that the scope of the concept of empathy is unclear. They question if it is mostly about affective mirroring, or about cognitive and emotional appraisal, embodied and behavioral assimilation, all of them, or if there more to how these processes occur [37]. These questions reflect that beyond establishing what is Empathy itself, the search among authors is for the essence, nature, and process of Empathy. Their objective is to observe if they are able to design a more natural, considerate, and long-term engagement with users through their own distinct perspectives [47]. Yet, when answering these questions, there is another concern that rises as current models of empathy being applied may not necessarily translate across projects, types of users, or for long-term relationship between a person and their technology [10]. To properly tackle these concerns, this article begins by offering a survey of works that speak to empathy and human-object relationship from across time and disciplines. The first objective of this paper are to introduce a strong theoretical concept as a proposed model that visualizes the process of realizing empathy, based on the ample analysis of the collected work in the survey. Secondly, the intended purpose of this proposed model, is to create an initial blueprint that may be applicable to a range of disciplines with clear must-have concepts important to consider for the realization of empathy between people and their technology. For this reason, after the model is explained, this paper exemplifies tools for its application and a couple of encouraging case study projects that begin to integrate this model into their interactive experiments.

Inspired on the array of works found, the model intends to leverage the strengths, parallels, and complimentary concepts found across the literature to present a new perspective as a set of stages-like process with key elements, and variables in order to reach Empathy and its results in a constant back and forth interaction. By viewing empathy not as a singular phenomenon, but as an active body of constant exchanges, can shed light on key attributes that may contribute to a sustained person engagement and experience with others, whether in building a relationship or as part of a collaborative scenario with technology.

## 2. Theories around Empathy

Empathy as a term has had great evolution and transformation since its origin during the late the 19th century and throughout many fields [25]. Its ephemeral yet tangible nature has intrigued scientists across multiple disciplines, including disciplines like HCI and Social Robotics, that hope to enhance or develop a model for an empathetic relationship between people and their technologies [43]. Despite numerous and on-going efforts to understand empathy, there are several main points in which most authors generally agree on: (1) It is a term with a no-universal definition; (2) There are sets of elements and variables that need to be present for an empathetic interaction to occur; (3) Because of the variables and different elements, it is a phenomenon difficult to pin point as a behavior and where it occurs in the brain activity; (4) There is a general consensus that empathy can occur at different depths and manners, from instinctual to reflective, from active to passive, cognitive to affective [46]; 5) Most agree that there is a clear pattern related to "imitation" or "mirroring" by the subjects of either specific embodied relatedness, cognitive reflection, or emotional circumstances [14]. This "mirroring" behavioral pattern has resulted in theorizing that predicting other's thoughts, actions, or emotions may play an important role in identifying whether empathy has occurred or not [3]. To expand on this further, in the next sections you will find the evolution of its definition, the impact of neuroscience, its presence in the arts, philosophy, and other disciplines, while also speaking to the relevant psychological evaluations around Appraisal Theory, Vicarious emotions in Empathy, Perspective taking or role-taking theories of empathy, and others.

## 2.1 Beginnings of Empathy

Empathy was initially born out of the concept coined as *Einfhlung*, by German philosopher Rober Vischer in the late 19<sup>th</sup> century, meaning "feeling into", describing the emotional experience of a person observing a piece of artwork they relate to, later extending to the embodiment felt with other types of objects and environments [25]. As a term, the origin is situated in the context of an expanding curiosity in psychological literature around sympathy, over how people perceive their feelings in nature, their contexts, and through the use of objects as tools of expression.

Theodore Lipps shifted the concept of Einfhlung beyond the relation with objects and art to how we are able to understand the mind of other people, by observing the instinctiveness of how people imitate and reflect each other's sensations and emotional expressions, equating it to the sense of relatedness people have when attracted to expressive artwork or structures [28]. Despite this jump forward, he did not infer differences between Einfhlung and sympathy. Though soon enough the blurred concepts between sympathy and empathy were evaluated and heatedly debated [28].

Even to this day sympathy and empathy are often confused due to their natured similarity and evidenced activity in the brain [9]. Yet as Darwall [17] assesses, the concepts are distinct. Philosophers and psychologists alike challenged Lipps' work hoping to clarify distinct processes in these terms, yet it was Edward Titchner in 1909 [17] whom by translating the concept into the term *Empathy* argued its difference from sympathy [23]. In essence, the argument is that sympathy relates more to the observation and care for oneself and others in a sense of compassion towards them from a third-point perspective, or what was termed *fellow-feeling*. While on the other hand, empathy has been described to be about embodying the emotions and the state of mind of the other as a tool for reflection and understanding, rather than for care about the other [17].

This observation grew in acceptance among scholars that quickly questioned if Empathy can lead to acts related to sympathy, like compassion. Theorists, among them philosophers and psychologists, followed this topic by testing when people are most empathetic with others, and observe how they react in those circumstances. These experiments since the 1960's, ranged within the idea that, first people may be most empathetic in highly emotional situation stemming from pain and misfortune, called vicarious emotion conditioning [9], and second that people observing others under these circumstances are generally more compelled to show compassion or aid them [32].

Correlated to the pain-sharing experimentations, some of these insights have led to theories that people are also able to cognitively identify the other's perspective, particularly exemplified in studies of perspective–taking in relation to Theory of Mind [4]. This means that perspective-taking and Theory of Mind may provide clues to how empathy occurs as well.

Though these concepts continue to be debated and inconsistently defined, the underlying differences are identifiable, stirring up the clear layering and complexity of the Empathetic experience [2].

## 2.2 Current Theories and Definitions on Empathy

### 2.2.1 Vicarious Emotions in Empathy and Appraisal Theory

Emotions and how they are perceived have played a big role in how Empathy is evaluated and defined. According to Wondra J.D., to "feel what other's feel" is a way of sharing affective states and experiences based on people's own memories or how they imagine would be their own experiences of what they perceive is other person's situation. These vicarious emotions can be achieved through different means whether it is mimicking, direct observation, assimilation, role taking, or memory recollection [46]. These tools facilitate the experience of these emotions, some more direct that others and while vicarious emotions are real, felt emotions by an observer, it is considered the imitation of the felt emotion the other is having. It is important to acknowledge that the observer also carries their own emotionality in the process. This will allow them to feel what the other is feeling, become a tool to act in compassion, or pivot away from the other [34].

But observing emotions or situations alone is not enough. Sensory, embodying inputs also add to this experiential revival as people identify with the other's experience. This can vary throughout the emotional spectrum, from physical pain, to general consequences of actions. This led to the observation that when people expressed their emotions or sensory experiences through their facial and body expressions, the observer also mimicked or vividly imagined that embodied feeling [31]. Therefore, when the observed person sees the observer mimicking their expression, as if they have also felt it, it is often assumed the other person is able to share those experiences. This does not mean that the people involved are experiencing the exact same emotion or perception of the circumstances. After all, everyone lives under unique conditions, genes, and personal perspectives. A concept that takes this into consideration is the *Theory of Appraisal in Empathy*.

It is believed that in the process of observation and imitation, the person appraises what they are observing rather than perceive it. To appraise, the person interprets the situation and emotions gaging what it means to them and how they believe they would experience it [41]. This inherently means that there is a cognitive, reflective layer to empathy that can be instinctive, as well as be a more profound thought and belief. This supports the fact that psychological theorists have been arguing, for a better part of the recent decades, that empathy more than relying solely on vicarious and personal emotionality, embodied mirroring, and visceral response, it can also be a process that requires a level of cognitive reflection, appraising people's situation [41].

This theory assumes that if a person appraises a situation, interpreting very closely to how the other is feeling and thinking, then Empathy would have occurred. To this end, the theory emphasizes on two important factors: that the observer has all information necessary given properly to correctly interpret the other with communicated emotions. If what is given is not enough, there might be confusion or misinformation causing the process of empathy to fail [46]. The second key element is that all emotions comes with a goal, in most cases a goal for well-being and understanding. This sense of having a goal is important, as it helps move the process of Empathy for those involved [41]. Whether the observer might be affected by the other's goal or not, it is a shared experience they can relate to. This entails that Appraisal theory helps to view Empathy more than just interpreting and deeply understanding the sentiment, embodiment, and knowledge of the other, but that is goes hand in hand with an emotional response, whether about care, show of compassion, expression of shared circumstance, or absolute rejection [5]. In any case, it continues to be a theory that assumes those involved are intrinsically motivated to understanding each other. It is clear that to elicit that vicarious response there has to be an innate sense of identity the actors identify with in the other [40]. This relatedness facilitates the imagining of the other's situation and appear empathetic, just because it is assumed, they know who the other is.

With a goal and function, empathy becomes a complex tool that can clearly be instinctive, but can also provoke more discerning reactions. It all depends on the goal for good state of well being for oneself and the other. During this moment of reflection, interpreting the other's situation and their own role, people weigh the risks, their morality, and the level of relatedness with the other.

Yet, empathy does not only occur when the people involved already identify with the other person. Empathy can also be evoked when people listen, read, and learn of other people [11] or despite the understanding, not have the willingness to have an empathetic reaction towards them. There are several ways Empathy can turn into an act of rejection or a push away from the observed. For example, if the other is too different than themselves, it becomes harder to understand and identify with them [20]. They can also reject if the defined concepts of morality, social rules and structures are affected and not within their taught definitions [32]. One other form of rejection can simply be that the observer, despite understanding and empathetic to the situation, deems any possible action too risky to their own well-being and personal goals [24]. As these forms of rejection do exist in empathy, this paper proposes a set of variables and elements that may facilitate a positive empathetic reaction, starting with simulating elements from the Attachment Model.

Some social psychology theories and practices believe that when provided flexible communication tools, a space that feels safe and inclusive, time for self-understanding and transform their intrinsic motivation, and allowed freedom to choose their role in the exchange, empathy can be practiced and guided [30]. An Attachment Model integrated into the guided practice is crucial to construct a caregiving system representative of Empathy [30]. One of such practices is *Perspective Taking Theory*. Independently of the reactions that may occur, when the subjects exercise observation and considers what the other is feeling, thinking, or acting, they are making a conscious effort to role-playing the situation of the other, even if in their own minds [18]. Though a more effortful process, previous studies have shown that with this guided practice to appraise the other's situation when the other is not someone the observer might initially relate to or to their situation, has provoked an increased compassion and reduction of bias [5]. This means that there can an active process of gathering information, listening, and reflecting on how to best interpret the other's point of view, independently of their differences.

In general, the current theories focus on how Empathy is a phenomenon related to emotion sharing, with the help of reflexive, embodied understanding coming to three main values of caring for others, understanding others, or validating other's emotions. Yet there is a lack of detail in most theories about the process itself of how the appraisal or perception identifies that empathy was created for the purpose of further engagement and in building a relationship [46]. In the search for clearer objectivity in this process in relation to creating a relationship between people and their technology, it might be worth to begin exploring how people make impressions of other people and objects.

Neuroscientific evidence shows that people are sensitive to social-cognitive demands when forming impressions on meaningful social objects, examining other's mental state and other traits [35]. This means that impression formation is not general but a selective process to further engage and have interest in the other. In contrast to inanimate objects, the impression formation in the brain does not activate in the same intensity as when forming impressions on other people, but it can be stronger when linked to a meaningful connection to them or as part of a story they can relate to [21]. To engage the impression, meaningful objects must carry relatable humanistic attributes or a strong assimilation to a person's values [1]. Understandably this impression formation can lead to further engagement and interaction with the object. This can be in any type of activity, but it might give further clues of how empathy can be provoked if part of the focus studied is on what are the tools to create engagement facilitating perspective taking. As seen throughout the previous theories, communicating information and emotions are important to convey. A format that has been used in scenarios of transforming conflict into moments of empathy and trust, is dialogue [26]. This is also explored and used as a way to have people get closer or interact with artificial agents, robots, and computers [10]. Dialogue is used as a way of connecting the engagement and naturally speaks to a process of desired understanding of the other emotionally and cognitively in order to drive trust and empathy [26].

*2.2.2 Neuro-scientific Presence of Empathy in the Brain*

The vicarious emotions in empathy and appraisal theory have been a big part of how scientists understand empathy in recent decades, mainly because the arguments are observed and detected in the human brain with the presence of mirror neuron and mentalizing system activity [13]. In order to detect the presence and trajectory of mirror neurons, which are considered a direct biological manifestation of empathy, and other activity related to feeling-sharing in the brain, fMRI is used as a main tool of reference.

      fMRI studies reveal that Empathy lives in several related behavioural regions of the brain. On one part you have the somatosensory, insula, and limbic regions of the brain where mirror neurons have been detected and where the activity related to perception, imitation, and embodied recollection occur; or how they imagine they would act and feel after observing another person's situation [36]. For example, Ashahr, Y.K., this region was also related to the concept of empathy distress as people observed others under duress. Yet, besides these regions of the brain, the desire for care almost automatically followed after observing distress. In this situation the septal region of brain activated, mostly related to pro-social behaviors, partner support, and trust [3]. Another fMRI study goes further than care, in a study that looked at indications of predicting prosocial behaviour. This study found that beyond the instinctual process of understanding the other's pain through observation and indicating motivation to care, that there are types of distress that require context. When people looked to the context of others, there was a layer of cognitive appraisal that helped interpret the other's situation driven by the mentalizing system. A manner of perspective taking, through their environment in order to observe and engage with more profoundness and care for that other [31]. The mentalizing system in the brain is composed by dorsomedial and medial prefrontal cortex, temporoparietal junction, posterior superior temporal sulcus and temporal poles [36].

These studies evidence three key points around empathy and its process: one that empathy may be induced by simply observing the other's emotional experience, but at times requires to actively learn about their context, meaning the environment they are in and what happened previously, taking their perspective to connect and understand them. This shows a direct correlation between mentalizing, mirroring, and empathetic altruism [37]. Second, this also demonstrates that empathy is physical, sensorial, and embodied. Taking perspective and instinctively mirror the other's emotions and feelings, moves the person to enact pro-social behavior, thus being part of the empathetic process [36]. Third, these studies reveal that empathy appears in the brain at different times and places to create a correct affective interpretation. It remains unclear the path through the different areas of the brain the empathetic process takes to support the changes that happen. This also means that an empathetic process is flexible and malleable depending on other factors like context, the level of relatedness the subjects have with each other, or the perspective adopted during the observation [43].

   These and many more studies related to mirror neuron activity and empathy have helped better understand key aspects of its process, providing tangibility to the array of definitions. But, as scientists continue to dig deeper, the understanding of the empathetic process becomes more complex and nuanced, therefore the exact mapping of where empathy happens becomes vague to pinpoint [36]. Though it is a natural phenomenon, the evidence shown points to the idea that empathy is not always an automatic response but can also be learned or practiced activity through time [44].

*2.2.3 A Philosophical tool for Empathy*

Philosophy takes a unique perspective that allows to see Empathy and other relational subjects and concepts from an analytical introspective point of view. The general observation around empathy has led philosophers to argue how it happens either purely by how people perceive someone else's emotions in a situation, or by how people perceive the other's situation overall. These arguments became a reference during a time when emotions and cognitive processes were studied individually, and continue to be part of a larger debate as neuro-scientific evidence supports aspects of both arguments [46]. Yet, part of the new evidence suggests that cognitive processes and emotionality occur simultaneously in the brain, analysis and emotional mimicry. This has led authors to evaluate what takes part in that cognitive analysis and emotional adoption, things like risk assessment, conversation, negotiation, and dialogue in order to maintain empathetic interaction [44]. David Bohm, a respected physicist and philosopher, introduced dialogue as a process tool for sustained social interaction, ideal to exchange thought, and expression.

Dialogue as a process is "aimed at the understanding of thought as well as exploring the problematic nature of a day to day relationship and communication with the goal of making a shared meaning" [11]. Dialogue is meant to be an open process where the people involved willingly express, share, and absorb mutual though without the push of trying to win or make any one point of view prevails. As previously observed from perspective-taking empathy theory, the meaning each person forms in their own minds in reflection and consideration of the other may be similar but never exactly the same. Yet identifying those differences, is the precisely the key to create something new in collaboration [11]. This kind of communication can only exist if people freely listen to the other without judgement and the desire to want to know them as who they are without intent of influencing the other. The actors involved must have an open mind-set interested in truth and coherence. Otherwise, without this premise, the communication will fail. Because the purpose is not to convert someone else's perspective to their own, but interpret meaning beyond their own assumption, it is suggested that people come into the dialogue willing to challenge those assumptions to then actively cooperate towards something new.

Beside the structure of openness while listening to the other without judgement or reserve, the space each person is given to interject and express themselves is also important in this process. This space should be perceived as safe and open enough for each person to have the sense of freedom to contribute their thought in the dialogue. This space may physical and time based [12].

As the space is provided for the people in the collective, the tool of communication, the language used in the dialogue, is also a collective variable to be considered [26]. Language is in great part a societal construct that is used to build individual assumptions, how society works, etc. This is why it is so important to pay attention in how things are communicated as individuals and as a collective. The type of language used depends on the collective context is being used in, whether its spoken, signing or other forms of communication.

As all these elements and structures unfold it becomes clearer that the process of a dialogue is cyclical with the potential for growth if the elements are taken with seriousness and care. As dialogues can shift and be dynamic in content and with how it is done, the consistency of practice can solidify the structure more and more having the actors involved feel more freedom and openness to delve deeper into how they make meaning [11]. This means that with time, people will be more willing to share content that is more personal and intimate.

From the very beginning as people look for truth and comprehension there should be a level of trust in the goal of a dialogue. This will translate to trusting each other as they continue to get to know each other in the process. With that time as people are more willing to freely share their personal and intimate perspectives, trust grows.

In parallel as trust grows, and consideration of being listened to is present, so is love, according to Bohm,D. What begins as a fellowship an initial care for collaboration, to active participation, and then friendship, means deep care or attachment can also grow out of a constant practice of dialogue [26]. Most importantly, this parallel growth is a sign that coherence is met consistently, motivating the people to continue the cyclical process [11].

This kind of dialogue is present in all our day to day relationships, including that with inanimate objects [33]. There is a similar way of communication with our objects as with others, an artist has a relationship with his tools and materials. Though artists might have a clear picture in their minds, the material speaks to their potential, resulting in something similar to the expectation. An artist should, listen to this material and what was created, and continue to evolve the work until it becomes a common result between the artist and their understanding of the material they used [22].

In summary, dialogue is a structured process with elements that need to be present in order to function like: space, language and clarity of openness for truth and coherence. Then these elements live within the premise of a constant conscious understanding of challenging personal assumptions and the willingness to cooperate with others. Once these elements and context are considered, only then can the steps of openly listening without judgement and reflective search for similarities and differences can be done in a constant cycle that makes constant meaning together. If this cycle continues and is practiced with time, the openness and freedom become a growth of trust and love.

As the process of a dialogue is present in all or most relationships as a tool for communication and interaction, it is only natural, based on case study works, that it may be an inspiring tool for creating empathy and trust [26]. As Empathy is not always an automatic response, the cyclical process of listening to the other in the search to match the other's perspective or assumptions, as a form of dialogue, may provide the necessary information needed to create Empathy. In consequence the practice of dialogue can elicit affective responses needed to best emulate an Empathetic interaction. As an example, during an initial exchange as you are welcomed into knowing one another, the actors pursue truth and comprehension in a dialogue. People are open to take the perspective of the other, searching to match the appraisal on what they feel, think and embody. So, when a person emits an emotional expression say a laugh, the other may reciprocate or mirror a response appraised to the situation. This initial indication allows for dialogue to open up, and for empathy to take place.

One of the literary works that has inspired this article, is the work by Seung Chan Lim, where in his book, he does a deeply–reflexive study of his own personal experience in coming to Empathize. His work, though not scientific in nature, has a comprehensive structure and terminology that resonates with current empathy work, the communication tool of dialogue, and the current applications of empathetic interaction with objects and technology [33].

*2.2.4 Empathy in the Arts and other fields*

From Economics and Game dynamics discipline, to anthropology and medical practices, empathy is present and models have been constructed to best serve these fields, each with interesting focuses from motivation, curiosity, to trust and drive [15]. To discuss in detail how these disciplines focus on these concepts is pertinent and relevant to the work, yet it goes beyond the immediate scope of this paper. This literature is considered in the manner in which the model presented is designed.

Despite the complexity and broadness of empathy definitions, there is something to explore when artists and designers continue to move and entrance people in sustained engagement through expressed tools of visual elements, music or dance. There is a long history, of how "empathic processes are essential to the aesthetic experiences of visual and other forms of art" [25]. These works create a story that the audience relates to as their own, mirroring their own perception and values. Their process to develop those stories may also provide clues as to how they are able to create empathy with an audience at a distance merely through the tool of expression. Most artists talk about having a dialogue with their materials while also reflecting deeply on their own vision. And with that reflection, embody their intention with the material they have at hand. Once created there are clear traces of motion, gesture, and intensity of emotion, where the observer can also interpret and embody themselves [22]. Brands also dig into understanding their audience, bringing them into the fold as part of their product and service development, so the perception of the brand continues to be relevant and intimate for each follower [6].

An influential literary work by Lim, recounts years of reflective experience around the process of making and how that relates to empathy. In his work, though not academic in nature, does echo much of what decades of study throughout a diverse communion of sectors have experienced in both observation and empirical evidence. The singularity of his observations is the clear and relatable metaphors and use of terminology to better understand beyond empathy itself, into what it means as a process. In principle he is constantly reflecting from the point of view of an artist and designer as if an intimate look inside his own process and of other artists.

With all the literature around art, aesthetic and its deep roots with empathy along with Lim's work here is a brief description of why it is important to consider this perspective of empathy. To decode the visual language is to decode a tool of communication or dialogue. To clarify, this language does have a grammar, a syntax, and series of formats that enable the optimal way of communicating with it. From the space it is viewed, the canvas it is drawn on, to the lines and use of color within its content, all have a role to play and a significance in its development [27]. And like any other art form, it requires time and practice, a constant iteration and conversation between the artist, materials and audience until they capture meaning between them, a significance that moves both the artist and the audience. In that dialogue, what the audience can see is the artist exploring all the different ways his tools allow him to wield it, like a carpenter knowing how to cut, bend, and join the wood to make it stronger. Through the iteration and practice, the artist learns to respect, listen, and care for his tools of expression the same as any professional that cares for their tools of work. It is of logic to these professionals, that the moment a tool is forced to act in ways it is not supposed to, the tool reacts in a negative way, as if it speaks to its owner on what it can and cannot do [33].

In this close observation of craftsmen, artists, and designers going through the intimate process of making, it becomes clear that with the right environment, context and intention, a meaningful empathetic relationship may occur for the general public as well with their objects. Even more so if they represent or serve a profound purpose in their lives. Here is where in this article it begins to become clear that it is only natural to assume that empathy and its process can occur with objects, and with more potential when they are technological smart objects.

## 3. Perspective on Applied concepts of Empathy in Human Interaction with technological objects

*3.1 Extended-Self Theory and the meaning of objects*

Extended-Self theory began with the meaning objects had to people for various reasons as they represented aspects of themselves that felt intimate and profound, in essence a part of themselves. Objects help support or express parts of a person's identity within a certain context, time frame, or aspect of their lives, which in turn, they develop a caring relationship with their objects, both ensuring their longevity, as needed, and mourning for their loss as well [7]. Since 1988 this theory continues to ring true and with even more relevance as people have gained digital, virtual, and smart possessions that deconstruct and are attributed a whole different set of needs from physical analogue objects [8]. As objects are able to identify their owner's patterns, overall behaviors and emotional cues, while also expressing on that data, the relationship people carry with their objects shift or rather becomes more intensified [42]. A whole set of relational variables that were present before, gain that much more importance like trust and freedom of control. For example: in [42]. users felt a better relationship with objects they perceived as their servants rather than an equal or when they exhibited relatable emotional cues. With analogue objects these existed, but as soon as the object no longer worked as it had, it may have often been kept around as keepsake or memorable trinket. Now these objects can talk back and some have human-like abilities like speaking with a naturalistic language or "remember". If these talking objects fail, they might not be treated with the same nostalgia as physical objects once did [39]. Many objects have reached a level of integration in people's lives that seem to have caused an empathetic interaction with their owners, for example, many Alexa users are already expressing love and deep connection with their device, indicating a possible empathetic interaction between them in the basic functions of the device. Yet there is still much to understand in the complexities of day to day interactions within a variety of contexts and needs a person may have.

*3.2 Quantified Self and empathetic objects:*

An area of knowledge that takes note of this is Quantified Self. This practice revolves around the premise of people being able to track their own habits, body statistics in order to improve their health and make their day to day habits more efficient [52]. This discipline has the objective to give people deep and varied knowledge of themselves so they can reflect and act in their own best interest. With a fast and rising following, its process and delivery is attractive as people engage in a conversation with their own bodies through data and the technological tools that provide that data. Embodiment continues to be a clear variable in how people perceive and feel other's experiences physically and emotionally, but there are times when people do not understand their own bodies or their habits impact creating dissonance between what they perceive, feel or should be feeling or embodying [51]. Quantified Self allows people to examine, learn, and above all, understand themselves and their situations. Though it is about self-knowledge, there is a sense of relationship building of their own bodies through their technologies that is highly emotional and reflexive as people hone their intuition around their habits and what they really need. The technological tools they use for these practices become essential to them, deepening their meaning and function beyond their core use as an extension of their emotional and physical support. Many practitioners delve deep into researching and understanding their tools for quantifying the habits and aspects of themselves that needs tracking, telling of the type of connection they seek of these smart objects [50]

The process in which quantified self practioners go through is very specific and particular in order to achieve self-knowledge with set markets that if not met can cascade into a derailed understanding of themselves. Their technological tools are not immune to this process and cn also confuse or skew the person using the device while tracking their activity. Yet the concepts derived from [53] do give an insight into the importance of listening and reflecting on information both quantitative and qualitative of themselves and how the tool captures them. Mainly because even though the device "listens" or captures data, the person also acquires a version of that knowledge with qualitative information: how they felt at each turn, the emotions they had, and their context that the device itself might not be able to contain. This duality means that the person must be flexible in understanding that the device might not be as complex as they need it to be while also reflecting on the importance raw data can have, better interpreting their situation and shift their perspective about themselves [48]. In essence this practice of constant listening to oneself and their surroundings while also having to reflect and analyse with that data, could be a way of empathizing with themselves and their tools.

*3.3 Persuasive technology\*\*\**

Many tools particularly used for quantified self situations tend to use an interaction intended to nudge or persuade a user towards a particular action [56]. This is part of the more general practice of persuasive technologies meant to serve as a way of supporting users in having a more efficient processes and habits [55]. In this sector the technology "listens" to the user, learns from their behaviors and sets goals with them.  By using persuasive techniques of interaction like game mechanics, it intends to steer the user in the direction they had set out for themselves. This active engagement with the user has caused a series of effects on the user's long-term interaction with the smart object's behaviour from over dependency to complete abandonment [39].

*3.4 Empathy and Prediction*

Practices like persuasion and nudging towards a set of actions implies that the technology is able to consistently predict the users needs without a motion to consider if the interpretation is correct or not. Even though, as a usability heuristic rule, the object provides tools for control and freedom of use, the bearing of suggested analysis with confidence tends to give assurance to the user that the techonolgy has done a good job of understanding the situation, at times even more so than the users themselves. This creates a sense of doubt that when verified can create a breach of trust in the object. And yet despite knowing that the technology is infallible, users continue to use them, starting an unhealthy relationship with their objects. This is why it considered important to address that making smart objects "predict" users needs does not necessarily equate to the objects empathizing with their users, but just that, they are trying to predict the users needs. This is not to say that people in general have patterns in their routines, way of thought and general behaviour, or that people would like for smart objects to facilitate decision making, but that predictability can be better integrated into the need for control and verification from the user in a more flexible and evident manner. I think this is something to be addressed as many believe that predictiveness and accuracy of predicting someone else's thoughts and actions are a result of being empathetic….yet this is not always the case. This concept is used a lot throughout commercial products and services.

*3.5 Empathy in Robotics:*

Given the strong role empathy plays in shaping communication and social relationships beyond others but also with themselves and their objects, it's is only fitting that empathy is also a major element in human-machine interaction [10]. There is evidence that people can feel, care for, or distress for their peers as well as game characters and even robots [3]. Paiva's survey work comprehensively addressed different case studies and agent-empathy models to best describe how robots and artificial agents can evoke empathy with their users. Despite the diverse definitions explored, the most comprehensive and used for agent or robot related empathy with users is De Waal's definition on empathy: the capacity to be affected by and share emotional states of another, assess the reason for another's state, and identify with the other, adopting his or her perspective." [59]. This definition has allowed to contextualize the study of empathy with robots be considered as a process in which that artificial agent or person have or express feelings congruent to the other's situation than with their own [60].

Although currently many artificial agents use the OCC (Cognitive Appraisal theory of Emotions, Ortony, Clore and Collins) model of emotions to be perceived as empathetic [61], Paiva, is able to complement their work with other studies and authors that also consider reflexive and dialogue driven models toward empathy. This also means that more and more these models result in designed interactions of empathetic driven and provoked by: vicarious emotions, a focus on care and distress empathy, dialogue driven interaction, and others [60].

To best achieve this behavior, Paiva (2017) considers that there are three main considerations: the mechanisms in which empathy arises, the modulation of emotions and intensity of empathy, and the empathetic response by which it is expressed and communicated for actions to be taken [60].

Part of the mechanism that has been widely evidenced is the instance of mimicry [45]. This imitation of another's expression means an openness to engage with the other, or that there is a level of relatedness. When a person sees an agent or object mimic or have human-like attributes and behaviors similar to their own expression, people will react to a mimicking that is similar or relates to that which was expressed by the agent [45]. Another, related aspect of mimicry is the anthropomorphic characteristics agents can be designed as a factor that may enhance social effects and emotions in humans. This is the initial impression people will have from these objects. With humanistic elements in both their aesthetic and behavior the observer will relate to them more and provoke meaning to them than when they are neutral [62].

Modulation factors considered are: features related to observed emotions, social relationships, and the context of the situation, as depending on the intensity it may impact more or less the perceived level of empathy. This means that behavioral transparency and consistency is important for the observer to be willing to further engage in empathy and rate the intensity of the empathetic interaction, this means elements of communication like language, expressiveness, movement, etc. This behavioral transparency must be seen as congruent with the information communicated to the person and the emotions linked to the content itself [44]. The existing relationship and known intention, that the agents might have with the target either when they know each other or through time may also impact the perceived intensity of empathy. It is not that same a familiar person or agent, than a new object. Finally, the context of the situation and its proximity also affect the level of intensity of empathy. When the actors engaged in empathetic interaction are in close proximity rather than absent or far, the empathetic experience will be felt more intense [12].

The last consideration of empathetic response, as De Vignemont & Singer argues, that empathy only exists of the observer is in an affective state and open to correctly appraise the other's situation without judgement. Therefore, the result of the empathetic process is a felt emotion by the observer which can trigger specific behavior. The response can be expressed or communicated through facial expression, body expression, physiological response, action tendencies or spoken [63].

In the case of case studies that evidence how artificial agents and robots can evoke empathetic interaction here are three.

One example was developed by Rosis et al. (2005) who created an Embodied Conversational Agent that acted as a therapist driven by a model of dialogue. The goal was to achieve involvement between the agent and the user to help change eating habits. The results correlated positively to the behavior change and the agent's involvement opening the analysis of its impact in evoking empathy [64]. In a similar case study called FearNot! The dialogue and emotion expression are used to motivate users in the interaction and response of engaging in empathic process. Other elements that played an important role was the context and clear goal of the agents and participants [65]. A case study more clearly exemplified with robots is Gonsior et al. (2012) who created a system where a robot called EDDIE was able to perceive facial expressions and engage in small-talk dialogue through identifying key words. In the use of this robot the experiment conducted had a control neutral expression while the experiment led the robot to be able to respond and shift its internal "emotional" state according to the situation and hat it perceived [66]. The result of this experiment showed that people felt more empathy towards the robot when it expressed emotions & mirrored the user's expression in contrast to neutral response. The use of small spoken dialogues with affective expression made the user want to help the robot in the exercise and exhibit empathetic behavior.

In all these case studies the three factors that needed to be considered were present some at a higher level than others, but when present provoked the empathetic response needed to achieve meaning between the actors.

Though there is still much more room for growth, already there are clear indications that introducing: clear signs of affective states mirrored from the observer's emotional state, use of dialogue (whether pre-determined or not) to provide information exchange as well as a tool to process consideration of the other, and finally the manner of response to that consideration are relevant in approaching an empathetic behavior with Robots.

### *3.6 Human-Centered Technology*

As we continue to observe the impact and value of giving a deep empathetic relation between people and their technological objects, it is even more so when they become active participators of our work and creative processes. Human Centered Technology is an umbrella term that engulfs many research lines that want to ensure people and their complexities are well integrated and considered within the design of technologies [67]. Another discipline beneath this term, can be Human-Computer Collaborations. Collaborating with technology is evolving as the complexities of human needs and reaction to their technological objects is revealed as well as the smart objects become smarter [68]. From the human perspective in its application, the interaction that they engage with the AI object varies with specific sets of behaviors that relates to the relevance of control and willingness to collaborate with the AI object [69]. What is evident across experimentation and studies is that AI objects and technologies will acquire enough knowledge to evaluate and decide on their own, but in the center of the collaboration is a dialogue of argumentation, whether between them or in the reflection the user makes on their own when implementing the technologies assessment [70]. To include the human needs within this collaboration means to consider their need for control and sense of freedom of choice, as long as that element is respected within the collaborative scenario with a smart object, they might perceive it as a peer, marrying the initial concept of how they relate to their objects as an extended form of the self. To acknowledge the vastness and limitations of each actor is t acknowledge they can compliment each other's knowledge and abilities. As much as technology can advance in relation to patterns and endless generation of possibilities, in every day circumstances, there are unforeseen variables that may occur and the machine will need the vision of the user in order to adjust their analysis. At the same, because of the efficiency, speed and data analysis the machines can do, the user may have the option to be willing and open to the machine's point of view. At this point they should have an open platform where they can argue each other's points for true collaboration [71]. Through this back and forth, there can be control, consensus and knowledge realization for both the user and the machine.

**4. Objective of Empathy and Summary of common elements**

Taking into consideration the evolution, transformation and degree of difference in the perception of Empathy across disciplines and their applications, the objective of Empathy also becomes spread out according to the needs and perspectives each take. Because the definition of Empathy is varied, to distinctly and ultimately define what is Empathy is beyond the scope of this paper. Rather, this paper accepts that Empathy can occur in multitude of scenarios with different kinds of subjects and levels of interaction. Taking into consideration the different perspectives across disciplines, rather than emphasizing on the instinctual or automated revelation of Empathy, this paper focuses on analysing how these different perspectives identify the process to reach Empathy with other people or with objects. The authors of this paper believe there are parallels that can be drawn among them, proposing a process with steps, elements, and variables that need to co-exist in order for the result in an empathetic interaction to occur. Based on the general acceptance of Empathy we will adopt the following definition of Empathy and its objective:

With this clarification of scope and definition related to this body of work, the following section exemplifies how different areas of knowledge apply their concepts of empathy and their key attributes to create or provoke sustained relational empathy.

## 5. A Novel Approach to Realizing Empathy

Based on previous literature, we envisage our empathy model as a living process where without some of the variables involved in it, it would not survive. It is a phenomenon that needs the right conditions in order to exist and even more variables to become consistent. It needs a rooted base, nourishment, a stable context, and a series of stages of interactions to be sustained. More precisely, we envision the terminology model from the works of Seung Chan Lim, Realizing Empathy: An Inquiry into the Meaning of Making [33] and David Bohm, On Dialogue [11]. These works were chosen because of the comprehensiveness they carry, while other works on Empathy, relationship with artificial agents, and reflective quantified self also draw similar parallels.

Most theories refer to a "relatedness", a mirroring of each other's experience, feelings, and thoughts, because they have in some way experienced the situation the other is in themselves. Yet, it is pivotal to understand that although people may relate to another person's situation, the metal self-reflection, feelings, and sensations are individual and unique to each person. A person will most likely never have the exact same feeling or knowledge that the other is feeling or knowing at any given time [47].

This sense of relatedness varies at different levels: perceived shared cognitive understanding, recognizing emotional states, and embodied reactions from physical situations [40]. As it can also vary in depth and comprehension. The profoundness of thought and emotional connection, as well as how complete or thorough the shared relatedness is.

How these levels, depths, and comprehensiveness are assimilated, can be immediate, like the instinct of riding a bike or getting burnt. Or it can be progressive, like two foreigners travelling in the same group tour. In this case other factors, such as their current context, bring these actors together other than the sentiment of a past shared experience. In this exchange, empathy may be realised if the actors are encouraged to a willingness and open attitude of continuously engaging in finding common understandings between them, similar to the process of a dialogue [11]. In this sense, to realize empathy does not always mean to feel, know, or experience exactly what the other has. Rather, it can also be about finding common ground with another, finding a way to respect and acknowledge the others needs independently of them agreeing or reflecting the same understanding over a topic [58].

*5.1 Attachment, Trust and Curiosity*

This recognized relatedness at all levels is an instinctive action of *attachment* to the other, i.e. a link that each actor decides how profound it might be [30]. Showing a willingness to explore that likeness between the actors is a result of *trust* with the each other [58]. The addition of *trust* and *attachment* are the base structure of empathy (Figure1). They are the roots that anchor those involved. Without both present, the content and variables that drive the process to realize empathy disintegrates the same way the dialogue process does [11]. The more the actors interact in empathy, the wider and more rooted trust and attachment develops between them (Figure 2) [15].

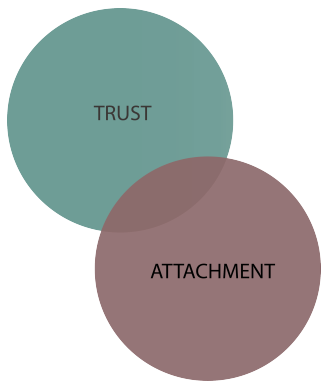

*Figure 1*

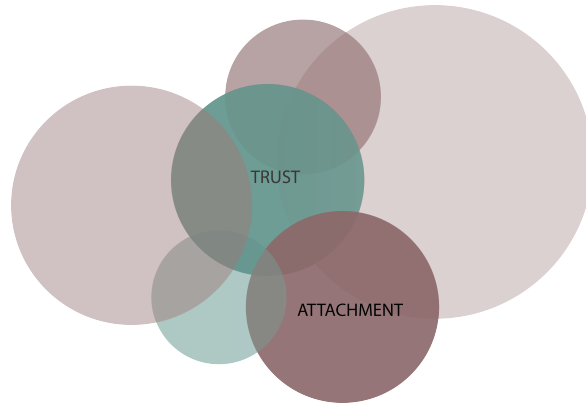

*Figure 2*

This marriage of trust and attachment exist intrinsically within those involved in the empathetic process. Despite their importance, attachment and trust alone are not enough to move them into realizing empathy. It must be driven by an internal motivation to interact with each other (Figure 4) [57]. This motivation can be interpreted as a constant cycle of curiosity, the *wheel of curiosity*, (Figure 3) a journey between being humbled and having courage: Humility by understanding that they do not know all the information of each other, and that they need to learn and cooperate from one another to truly understand the other. Courage, because with humility and willingness to cooperate, comes the active searching for that knowledge.

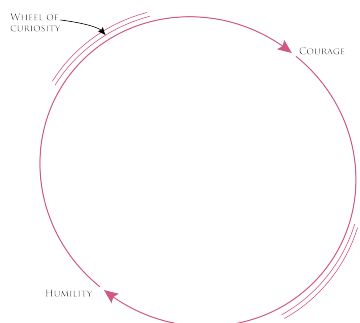

*Figure 3*

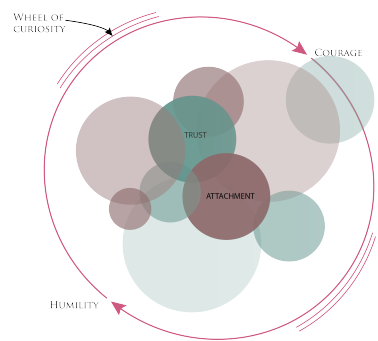

*Figure 4*

Adopting Seung Chan-Lim's terminology [33] those involved put aside their pre-conceived notions and are willing to delve and learn from a different perspective

[58]. Not all actors are the same in their perception or extent of humility and courage. There is no right level of humility or courage, nor is it something this paper seeks to find. But we can define a set of stabilising elements for the *wheel of curiosity* to remain consistent.

*5.2 Stable elements*

In order to make the process of realizing empathy actionable, it needs a set of elements that stabilize the exchange (figure 5). These are consistent components allowing for a system to flourish:

*5.2.1 Environment*

The space or environment depends on the context of the interactions and the actors involved. The space must be providing a sense of freedom and dynamism, open and intimate enough proportionate to the situation as the information travels directly between those involved. If the space between the actors is to wide, the message may get lost or not arrive fully, while if the space between is too close in relation to the context, the personal boundaries of those involved might feel violated, restricting the incentive for curiosity.

*5.2.2 Communication*

 A common tool of communication between the involved agreed upon, formally or not, to get the message across, whether it is a spoken language, metaphors, body movement, art, music, or stimulating elements. If those involved are not willing to adapt, accept, or agree on a common language the communication fails or may be misinterpreted.

*5.2.3 Open Mindset*

An open mindset requires for those involved to consciously position their mental state to practice an open behavior. This open behavior allows the other to interact with them accepting being the receiver of what they want to express. This does not mean that the person must agree or accept as their own the others viewpoint, it merely means that they are welcoming new information for them to consider.

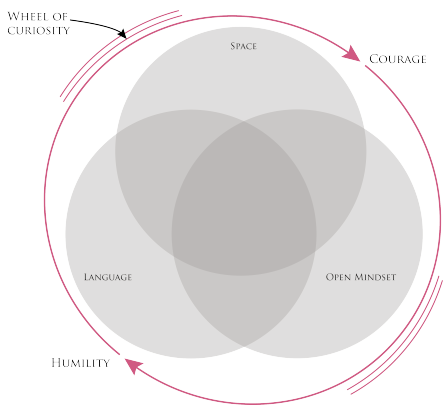

*Figure 5*

*5.3 The Process*

The following steps (Figure 6) enact the realization of empathy in the following way:

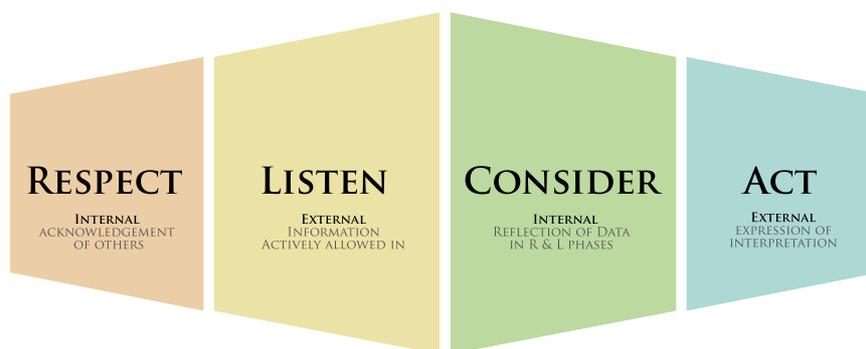

*Figure 6*

*5.3.1 RESPECT*

Those involved in the process acknowledging that the others are who they are [?], without judgment or wanting them to change. No matter the context, or what will be said, they must engage knowing that they will remain as who they are.

*5.3.2 LISTEN*

Refers to actively and widely collecting what the other is expressing externally. This means that the mind of the receiver must not hold pre-conceived notions, judgments, or comments at hand. Listening does not mean to silently wait until the other finishes their turn to express themselves; it means to take in the message as the raw data for comprehension and generating meaning among them. This is not an easy step; it takes conscientious practice, particularly when the meaning does not come as naturally as the actors would believe. Listening is being with them, attentively, and openly being a repository of their expression.

*5.3.3 CONSIDER*

Describes the internal process when actors internalize, process, or actively reflect on the message and information collected both through the respect and listening phase. It is the cognitive, emotional and embodied trial and error occurring inside the mind of each actor before they respond. It is the moment in which the actors choose tools to use in order to examine if their interpretation resembles what the other is trying to express. It is the moment that they test if the information they reflect upon becomes knowledge.

*5.3.4 ACT*

Act would be the last step as the actors externalize their interpretation in the hopes that the other accepts this as reminiscent of their own message. This expression towards bringing them closer to the others perspective is done accordingly to consideration they have had of both what they have heard and the extent at which they respect the other.

When the result of acting or responding to the other does not have the likeness expected, then the wheel of curiosity has the job to fuel the process back up again (Figure 7).

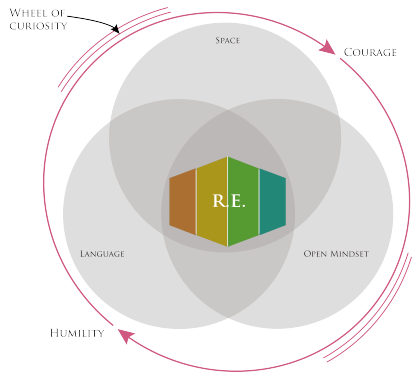

Figure 7

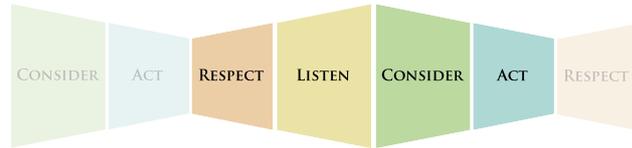

Figure 8

As represented in Figure 8, this process continues iterating by widening and focusing again and again, until a series of meaningful insights string between the participants. This activity that opens out and back in is like a wave, as each previous step informs and adds to the following wave moving information continuously.

Empathy can appear and expand as well as it can partially disappear or dim. Acting like a wave, as it opens up to the individuals interacting, and then closes as meaning is created between them, as so on. It is a wave that may shift in size and acuteness, but never truly, completely disappear. While if there is continuity in the process it may help knowledge, trust, and attachment expand, making the relationship grow.

*5.3.5 The Wave of Realizing Empathy*

When the result of acting or responding to the other does not have the likeness expected, then the wheel of curiosity has the job to fuel the process back up again. This process continues after by widening and focusing again and again, until a series of meaningful insights string between the participants.  This wave that opens out and back in is like a wave, as each previous step informs and adds to the following wave.

In this respect, the waves and each step may vary in size and breadth determined by the time and depth needed to execute them (Figure 9). There is no set amount of time or limit of range in each wave. This means that the steps may even interlace with each other as the actors make sense of the information and the other's needs.

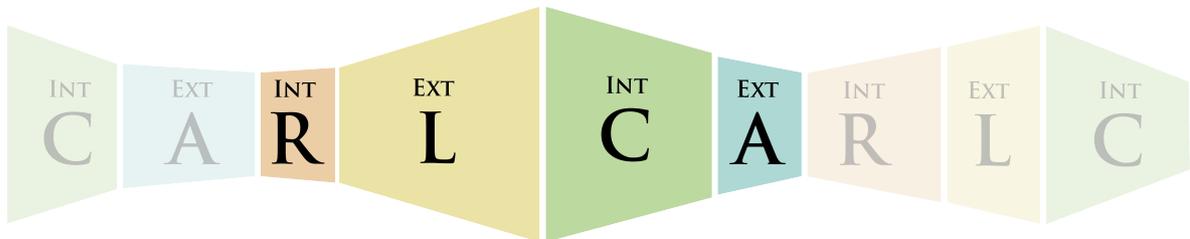

*Figure 9*

### 5.3.6 Meaning

The objective of the process of realizing empathy is to provoke meaning between the actors within the process. This meaning is ephemeral, living in the space between them in that moment and in each of them in their own interpretation that best assimilates to others, but never precisely the same. A similar phenomenon occurs to our perception of past events when stored in our memory as we build metaphors and elements that support our perception, which differs from other people's perception of those events even if both were present in that exact moment. Yet the reality of that event, though those involved may agree, the subtleties of what happened is perceived differently in each individual as they take on different perspectives. Therefore, the raw nature of meaning only exists in the time and space in which it occurred, the internal meaning is the insight that transforms the actor's own personal beliefs and further actions their behavior within their contexts.

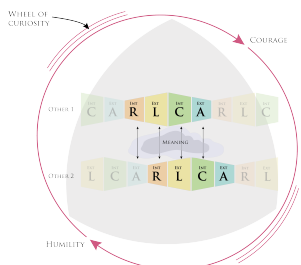

*Figure 10*

*5.3.7 Relationship Growth*

As the waves in the process of Realizing empathy continue to generate and add meanings and insights, the roots of attachment and trust continue to grow and expand. The consequence of this growth is a relationship being built; making the process of realizing empathy the link of the multitude of interactions that becomes a relationship. The longevity of that expansion is only relevant to the context of the relationship, the actors involved and the purpose of both (Figure 11).

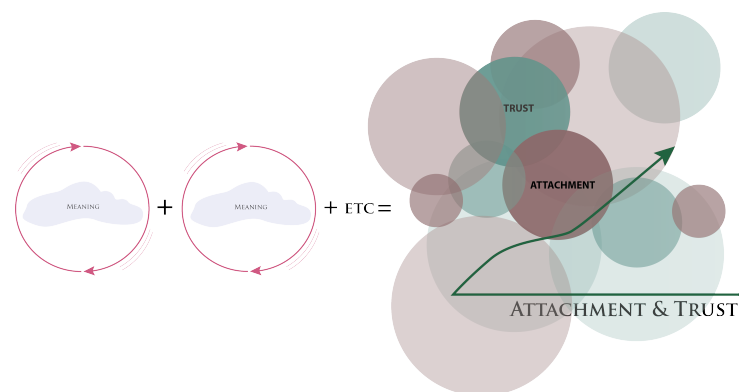

*Figure 11*

**6. Translatable applications and Case Studies**

Each layer and element mentioned can and may be translated into tangible example for smart objects and robots.

Design and Artistic elements for language, interaction for space and embodying tools, psychological tools to reinforce instruction and respect, while philosophical approaches allow for curiosity and open mind-set.

*6.1 RELATEDNESS, ATTACHMENT AND TRUST*

Whether engaging with other people or objects, the person in the scenario must have an understanding of what links them to that object or person, what is their purpose for engagement. Without a clear understanding of what is the role of each actor, the interaction becomes ambiguous and unsustainable. Once established, the role of that other person or object can shift, evolve or transform depending on how those interactions unfold nurturing a relationship.

As previously stated, this decision to explore the relatedness needs to be intrinsic for that attachment and initial perceived trust to take place. This means that when introduced to an object the engagement must be voluntary, offered but not provided, it must be felt like a choice unless the person already accepts that through previous knowledge that it is in their best interest to engage with this object. Freedom, sense of control and particularly a sense of security are of outmost importance for attachment to come about.

These feelings of: freedom, control, and security can be established in a scenario similar to that of a commercial store, where

How the different elements translate into tangible elements of behavior. By mimicking human behavior and culturally significant cues attached to affective and cognitive information. For example, using certain movements and pattern of sounds may indicate a state of information. If that movement and sound attached to color behavior of LEDS that meaning may shift from affective to informative.

These elements become the language used. The space would be determined by the context of use by the person in contact with the object and the function it is given. The attitude the object has depends if it is given nudging, persuasive qualities. If there is little to none persuasive qualities, the attitude may be deemed open minded, vs. a highly gamified experience with stated guidance, the object's attitude is clear and closed.

## 6.2 Current User Case works related to this model

Current future scenarios (social robots, health, artificial intelligence environments for human-system collaboration.)

When introducing this concept model into the social robotics sector, there is a concern that we hope to contribute towards a solution. The problem we see is the moment people do not understand what the robot is for nor for whom, and as the exploration continues this confusion tends to provoke the viewer or person it interacts with the robot to abuse of it or push a kind of function that is not intended for to do. These forced functions can range from actual physical activity as well as cognitive/affective states. This means that the key aspects that are dictated within the empathetic model mentioned may provide clues as to initiate and maintain a sense of ownership and connection between the object and person of interaction. If there is a sense of ownership and sense of what the object is meant to provide them, the actions that come after relating to that sensibility.

The model that we present, particularly, highlights the importance of voluntary, intrinsic motivation to participate, engage, with the understanding of what the object is. There must be a phase where the information of who or what they are approaching is meant for. Whether it is meant to be specific kind of companion with certain abilities or to have specific functionalities. Social robots like Aiboo, make very clear by its physical design that they are meant to be your pet, a companion that acts like a pet. Other novel robots like Pepper, provide a very broad introduction to those that interact with her making a relationship with people limited and with no clear direction.

This also goes for people to the object. People need to have a sense of their role, what they are meant for in a relationship in order to begin and continue in one. Their primary role in objects like robots maybe to be the taught, in case of educational robots, the priority of survival in the case of war robots, the patient, the caregiver, etc. Again, in more ample robots with wide sense of role, the person's role also feels broad with no significance to the life or function of the robot.

## 6. Conclusions

The collection of interactions must be with a bi-directional intention and exchange. The process to realize empathy in a dialogue format, may provide help to encounter meaning but also allows for evolution beyond the initial goal met. Products and services have a life before and after the encounters with them. Needs always evolve, as long as it is clear that the empathetic object accompanies them always in the quest of supporting them in resolving those needs, whether constant or not, the life and companionship between them is lasting.

**Acknowledgement**

We thank our research group, Grup de Recerca en Tecnologies Mèdia, in particular participants of the Seamless Interaction Group: Sergi Navarro, Gerard Serra, Flor Santianello and others. Marialejandra García Corretjer thanks the support of the European Social Fund (ESF) and the Catalan Government (SUR/DEC) for the pre-doctoral FI grant No. 2016FI_B1 00080. This work has been partially funded by SUR/DEC (grant ref. 2014-SGR- 0590).